\documentclass[]{jfm}

\usepackage{graphicx}
\usepackage{newtxtext}
\usepackage{newtxmath}
\usepackage{natbib}
\usepackage{hyperref}

\usepackage{amsmath, bm}
\usepackage{amsfonts,xparse}
\usepackage{amssymb}
\usepackage{overpic}
\usepackage{rotating,multirow,bigdelim,upmath}
\usepackage{tikz}
\usetikzlibrary{calc}
\usepackage{amsmath}
\usepackage{epstopdf,color}
\usepackage{mathtools}

\usepackage{graphicx}
\usepackage{newtxtext}
\usepackage{newtxmath}
\usepackage{natbib}
\usepackage{hyperref}

\hypersetup{
	colorlinks = true,
	urlcolor   = blue,
	citecolor  = black,
}

\newcommand{\RomanNumeralCaps}[1]
\linenumbers

\title{The interpretation of the amplitude modulation coefficient and a transport-based alternative}

\author{G. Cui \and
	I. Jacobi \corresp{\email{ijacobi@technion.ac.il}}}

\affiliation{\aff{1}Faculty of Aerospace Engineering, Technion Israel Institute of Technology, Haifa, 32000, Israel
}

\begin{document}
	\maketitle
	
	\begin{abstract}
		The amplitude modulation coefficient, $R$, that is widely used to characterize non-linear interactions between large- and small-scale motions in wall-bounded turbulence is not actually compatible with detecting the convective non-linearity of the Navier-Stokes equations. Through a spectral decomposition of $R$ and a simplified model of triadic convective interactions, we show that $R$ actually suppresses the signature of convective scale interactions, and we suggest that what $R$ likely measures is linear interactions between large-scale motions and the background mean flow. We propose an alternative coefficient which is specifically designed for the detection of convective non-linearities, and we show how this new coefficient, $R_T$, also quantifies the turbulent kinetic energy transport involved in turbulent scale interactions.
	\end{abstract}
	
	
	\section{Introduction}
	\label{sec:intro}
	
	\citet{Hutchins2007} first reported amplitude modulation (AM)-type behavior  between large- and small-scale filtered signals in wall-bounded turbulence. They decomposed the fluctuating streamwise velocity, $u$, into a large scale signal, $u_L$ and a small scale, $u_S$, and noted that large variations in the large-scale tended to correspond to changes in the envelope of the small scales.  \cite{Mathis2009} then introduced a correlation coefficient, $R(y)$, to quantify this AM as a function of wall-normal distance, $y$, by defining the large-scale filtered envelope of small scale fluctuations, $\mathcal{E}(u_S)$, and then calculating:
	\begin{equation}
		R(y) = \frac{\left\langle u_L  \mathcal{E}(u_S) \right\rangle }{\sqrt{\left\langle u_L^2 \right\rangle} \sqrt{\left\langle \mathcal{E}(u_S)^2 \right\rangle \vphantom{\left\langle u_L^2 \right\rangle}}}
		\label{eq:R}
	\end{equation}
	Their AM coefficient was based on the cross-correlation analysis developed by \cite{Bandyopadhyay1984}, but provided a simpler way to observe the variation in AM across the wall region. \cite{Mathis2009} noted that the profile of $R(y)$ appeared surprisingly similar to the profile of streamwise skewness, and \cite{Schlatter2010} subsequently demonstrated that the AM coefficient is not independent of the skewness. \cite{Mathis2011a} then showed how $R(y)$ is analogous to one of the cross-terms found inside a scale-decomposed skewness. Since then, the coefficient has been widely used as a diagnostic for quantifying scale interactions. 
	
	Amplitude modulation, in the sense of \eqref{eq:R}, was explained by \cite{Mathis2009} to mean that the large-scale velocity signal, $u_L$, modulates some small-scale carrier signal to produce the observed small scale signal, $u_S$. Thus $R$ should detect evidence for quadratic interactions between the large- and small-scales in the velocity signal, $u$. \cite{Duvvuri2015} rewrote $R$ in spectral form as a summation of real-valued Fourier modes over a region of wavenumber space corresponding to large- and small-scales that are triadically related. They showed that $R$ does, in fact, measure the energy associated with large- and small-scale velocity triads.
	
	But, the fact that three velocity modes are triadically linked does not mean that they are the result of a non-linear interaction. The standard way to establish that members of a triad are actually the result of a non-linear interaction is to show that the phases of the modes are also consistent with the non-linearity \citep{Kim1979}. But, as \cite{Duvvuri2015} derived, the phase information in $R$ appears only as a weighting factor, such that triads associated with quadratic non-linear interactions are weighted more than triads without. But the resulting value of $R$ does not indicate whether it is dominated by non-linear, AM behavior or just energy distributed spontaneously in wavenumber triads.
	
	More problematically, the particular phase weighting factor that is built into the definition of $R$ captures pure quadratic interactions between velocity modes, of the form $\mathbf{u} \cdot \mathbf{u}$, as opposed to the convective non-linear interactions between velocity modes,  $\mathbf{u} \cdot \nabla \mathbf{u}$, that are anticipated from the Navier Stokes equations (NSE). Indeed, it will be shown that the weighting factor inherent in $R$ actually suppresses the signature of convective non-linearities in turbulence, and thus cannot measure the presence of turbulent scale interactions. This is consistent with recent work by \cite{Andreolli2023} questioning whether perceived amplitude modulation behavior is actually associated with triadic scale interactions.
	
	In this study, we decompose the AM coefficient, $R$, based on the biphase, $\beta$, show how the definition of $R$ actually excludes the convective non-linearities that are responsible for inter-scale energy exchange, and instead likely represents linear interactions with the mean flow. We then propose a modified coefficient, $R_T$, that is compatible  with detecting convective scale interactions, and can also be interpreted naturally in terms of turbulent kinetic energy transport.
	
	\section{The Interpretation of $R$}
	
	\subsection{Bispectral Decomposition of $R$}
	
	\cite{Duvvuri2015} showed that the AM coefficient, $R$, can be expressed as a sum of purely triadic modal energies. However, they utilized sine functions as their Fourier basis, which slightly obscures the true phase-weighting embedded in $R$. Therefore, we begin by rewriting $R$ in terms of complex Fourier modes, after which we will examine how to interpret the weighting. The streamwise velocity fluctuation, $u$, can be written as a Fourier series over streamwise wavenumber, $k$, with time-dependent, complex-valued Fourier coefficients, $\hat{u}(k,t)$, according to:
	\begin{align}
		u(x,t) = \sum_{\mathclap{\substack{\forall k \, |\\ |k|<\infty}}} \; \hat{u}(k,t) e^{ikx}. \label{eq:fourier}
	\end{align}
	Then, following the procedure in \cite{Mathis2009}, the low-pass filtered large-scale signal, $u_L(x)$, and the remainder signal, $u_R(x)= u(x,t)-u_L(x,t)$, can be written in terms of the filter cutoff wavenumber, $k_f$, as:
	\begin{align}
		u_L(x,t) = \sum_{\mathclap{\substack{\forall k \, | \, k_f > |k|}}} \; \hat{u}(k,t) e^{ikx}, \qquad
		u_R(x,t) =\sum_{\mathclap{\substack{\forall k \, | \, k_f < |k|<\infty}}}\; \hat{u}(k,t) e^{i k x}
	\end{align}
	We employ the simple quadratic envelope \citep{Jacobi2013}, to define the magnitude of the small-scale fluctuations as $u_R^2(x,t)$. Then filtering with the same low-pass filter above, we obtain an envelope signal:
	\begin{align}
		\mathscr{E}(x,t)=\sum_{\mathclap{\substack{ \forall k', k'' \, | \\|k'|, |k''|>k_f \\ |k'+k''|< k_f}}}\; \hat{u}\left(k',t\right) \hat{u}\left(k'',t\right) e^{i\left(k'+k''\right) x}
	\end{align}
	The spectral definitions for $u_L$ and $\mathscr{E}$ can then be substituted into \eqref{eq:R}. Ensemble averaging, denoted $\langle \cdot \rangle$, with the assumption of stationarity then yields:
	\begin{align}
		R(y) &= \frac{1}{\Omega} \sum_{\substack{\forall k \, |\\ \left| k \right|<k_f } } \qquad \sum_{\mathclap{\substack{ \forall k', k'' \, |  \\  |k'|, |k''| > k_f \\ k' + k'' = -k}}}\; \text{Re} \left\{ \left\langle  \hat{u}(k') \hat{u}(k'') \hat{u}^*(k'+k'') \right\rangle \right\} \label{eq:RR}
	\end{align}
	where the normalization factor is defined as $\Omega = \sqrt{ \vphantom{\left\langle 	\mathscr{E}(x)^2\right\rangle } \left\langle u_L(x,t)^2\right\rangle } \, \sqrt{\left\langle 	\mathscr{E}(x,t)^2\right\rangle }$, and the real part is denoted $\text{Re}\{ \cdot \}$. We have written $R$ explicitly in terms of a double sum in wavenumber space, first over all the individual small scales, $(k',k'')$, that form triads with the large scale, $k$, and then over all of the large scales, $k$, that are within the wavenumber filter cutoff, $k_f$. This result is consistent with \cite{Duvvuri2015}, except for the use of complex Fourier basis functions.
	
	We note that the bispectrum for the velocity signal is defined as $B(k',k'') =  \left\langle  \hat{u}(k') \hat{u}(k'') \hat{u}^*(k'+k'') \right\rangle$, and thus $R$ is just the real part of the bispectrum summed over a range of wavenumbers that demarcate the triadic relation between two small scales, $k',k''$ and a large scale, $k$. In terms of the magnitude $|B|$ and phase $\beta$ of the complex bispectrum, $R$ is given by:
	\begin{align}
		R(y) &= \frac{1}{\Omega} \sum_{\substack{\forall k \, |\\ \left| k \right|<k_f } } \qquad \sum_{\mathclap{\substack{ \forall k', k'' \, |  \\  |k'|, |k''| > k_f \\ k' + k'' = -k}}}\; |B(k',k'')| \cos{\left[\beta(k',k'') \right]} \label{eq:Rb}
	\end{align}
	The sum of the bispectrum over all wavenumbers, when normalized, is just the skewness of the velocity signal \citep{Kim1979}, and therefore this partial sum is also consistent with the decomposition of the skewness performed in \cite{Mathis2011a}, where it was shown that $R$ constitutes the part of the total skewness associated with large- and small-scale modes. 
	
	In order to identify what is actually being measured by $R$, we need to interpret the bispectrum magnitude and biphase. The bispectrum is often described as a measure of the energy density associated with non-linear, triadic interactions, but in order to see what it represents in the context of turbulence, we develop a simplified model problem based on the NSE and calculate $B$ and $\beta$. Then we relate the scale interactions from the NSE to the value of $R$.

	\subsection{Convective Triadic Interaction Model Problem}
	
	Consider a unidirectional, instantaneous velocity signal that contains a large-scale, streamwise velocity mode with streamwise wavenumber, $k$, and phase-speed, $c_k$. We are interested in modeling the interactions between this large-scale mode and other modes in the flow, including the mean. The relative mean flow felt by this large scale is just the difference between the local mean velocity, $\overline{u}$, and the phase speed of the mode itself. Since the NSE are Galilean invariant, we can simply shift to the moving frame of the large-scale mode. Then the large-scale mode will exhibit a velocity discrepancy with the mean flow: very near the wall, where large-scale motions (LSMs) tend to advect faster than the mean, $\overline{u} - c_k < 0$, and far away from the wall, where the large scales tend to advent at speeds slower than the mean, $\overline{u} - c_k > 0$, as reported by \cite{DelAlamo2009}. This Galilean shift means that the DC component of the Fourier-transformed instantaneous velocity signal is not zero, $\hat{u}(k=0,t) = \overline{u} - c_k$, and depends on wall normal location. 
	
	Considering the simplest possible case of mean and fluctuating interactions, we assume that the large-scale mode, $k$, is involved in two triadic interactions: one linear interaction with the mean flow, i.e. with a triad containing the zero wavenumber $(0,k,-k)$; and one non-linear interaction with two fluctuating velocity modes, given by the triad $(k',k,k'')$. We assume that each interaction is governed by the convective term in the instantaneous NSE, written in spectral form as:
	\begin{align}
		\frac{d}{d t}\hat{u}(-p, t) & =i p \sum_{\mathclap{m+n=-p}}\; \hat{u}\left(m, t\right) \hat{u}\left(n, t\right)
	\end{align}
	for a general triad $(m,n,p)$ where $m+n=-p$. (For the unidirectional flow case, the pressure term drops out of the spectral NSE, and we neglect the viscous term for simplicity, although we will comment on its effect below.) Each complex Fourier coefficient can be written in terms of a magnitude and phase as $\hat{u}(p,t) = |\hat{u}(p,t)|e^{i \phi_p}$ where the phases $\phi_p$ are assumed random. Then we substitute the two interactions of interest to obtain the coupled system:
	\begin{align}
		\frac{\partial}{\partial t}\hat{u}(k, t) & =-i k \, \hat{u}\left(0, t\right) \hat{u}\left(k, t\right)\\
		\frac{\partial}{\partial t}\hat{u}(-k'', t) & =i k'' \, \hat{u}\left(k', t\right) \hat{u}\left(k, t\right).
	\end{align}
	We want to describe the Fourier coefficients after the interaction, $\hat{u}(p,t+\Delta t)$, in terms of the inputs to the interaction, $\hat{u}(p,t)$, so we linearize the time derivative for each wavenumber:
	\begin{align}
		\hat{u}(k, t+\Delta t) & = \hat{u}(k, t) -i k \Delta t \, \hat{u}\left(0, t\right) \hat{u}\left(k, t\right)\\
		\hat{u}(-k'', t+\Delta t) & =\hat{u}(-k'', t) + i k'' \Delta t\, \hat{u}\left(k', t\right) \hat{u}\left(k, t\right)\\
		\hat{u}(k', t+\Delta t) & = \hat{u}(k', t)
	\end{align}
	where $\Delta t$ represents the interaction time between the large-scale and other modes in the system.	We also assume that the $k'$ component is unchanged with time, as it is only an input to one of the interactions. Finally, we write the post-interaction complex Fourier coefficients: 
	\begin{align}
		\hat{u}(k, t+\Delta t) &= |\hat{u}(k, t)| \sqrt{1+k^2 \Delta t^2 \hat{u}(0, t)^2}  e^{i(\phi_{k} + \phi_0)} \\
		\hat{u}(-k'', t+\Delta t) &= |\hat{u}(k'', t)|e^{-i \phi_{k''}} + k'' \Delta t |\hat{u}\left(k', t\right) ||\hat{u}\left(k, t\right)| e^{i( \pi/2 + \phi_{k'} + \phi_{k})}\\
		\hat{u}(k', t+\Delta t)  &= |\hat{u}(k', t)| e^{i\phi_{k'}} 
	\end{align}
	where the additional phase contribution associated with the mean flow interaction is given by $\phi_0 = \tan^{-1}{\left[-k \Delta t \, \hat{u}(0, t) \right]}$.
	
	The phase, $\phi_0$, represents the strength of the modal interaction with the mean flow. Because this phase will be important in the subsequent analysis, we briefly consider its limiting values by considering the argument of $\phi_0$ as a ratio of timescales. Recall from above that $\hat{u}(0, t)$ represents the difference between the mean velocity and the convective velocity of the large-scale, $\overline{u}-c_k$. Then, $\Delta t_0 = -1/k \hat{u}(0, t)$ is the time scale of the large-scale interaction with the background mean flow. The other time scale, $\Delta t$, came from the linearization, and we interpreted it to represent the interaction time between the large-scale and other modes. In terms of these two time scales, $\phi_0 = \tan^{-1}{\left[\Delta t / \Delta t_0 \right]}$. When the mean flow interaction is dominant (i.e. its interaction is very rapid): with $\overline{u}-c_k < 0$, near the wall, $\phi_0 \to +\pi/2$; for $\overline{u}-c_k > 0$, far from the wall, $\phi_0 \to -\pi/2$. And when the mean flow interaction is weak and the scale interactions are dominant (and thus very rapid), then $\phi_0 \to 0$. 
	
	(Including the effect of kinematic viscosity, $\nu$, and defining a viscous time-scale as $\Delta t_\nu = 1/k^2 \nu$, we can write a more general expression for the phase shift, $\phi_0 = \tan^{-1}{\left[\Delta t / \Delta t_0 \left(1- \Delta t/\Delta t_\nu \right)^{-1} \right]}$. Therefore, when the viscous time scale is similar in magnitude to the interaction time scale, i.e. when viscosity is dominant and the viscous time-scale is relatively short, then $\phi_0 \to \pm \pi/2$ and the viscous effects simply amplify the effect of the mean flow interaction.)
	
	Having calculated the spectral energies for the three triadic components in the instantaneous velocity signal $u$, we calculate the bispectrum and biphase in order to interpret $R$ for this model problem. The bispectrum is given by
	\begin{align}
		B(k', k'' ; t+\Delta t) 
		&= \left\langle 	\hat{u}(k', t+\Delta t) 	\hat{u}^*(-k'', t+\Delta t) 	\hat{u}(k, t+\Delta t)  \right\rangle.
	\end{align}
 Substituting (and dropping the explicit time notation), and ensemble averaging, yields
	\begin{equation}
		B(k', k'') =  k'' \Delta t \sqrt{1+(k'+k'')^2 \Delta t^2 |\hat{u}(0)|^2} \left\langle |\hat{u}(k')|^2 |\hat{u}(k'+k'')|^2 	 \right\rangle e^{i( -\pi/2 + \phi_0)}, \label{eq:toy}
	\end{equation}
	and the biphase $\beta =  -\pi/2 + \phi_0$. Therefore, we see that in the absence of mean flow interactions, the biphase for the convective non-linearity of turbulence is $-\pi/2$. (More generally, it is $\pm \pi/2$, but this model problem considered the triad $k'+k''=-k$ and not $k'+k''=k$.) The $\phi_0$ contribution to biphase appears only as a result of a coupled mean flow interaction and does not appear due to simply including additional triadic interactions.
	
	Now we use the bispectrum from the model problem to examine how the nonlinear convective interactions of turbulence influence the value of the AM coefficient, $R$.
	
	\subsection{Model $R$ for Pure Convective Scale Interactions}
	
	In the limit of pure convective interactions between fluctuating scales with no mean interaction, the bispectrum in \eqref{eq:toy} can be simplified to
	\begin{equation}
		B(k', k'') \approx  k'' \Delta t \left\langle |\hat{u}(k')|^2 |\hat{u}(k'+k'')|^2 	 \right\rangle e^{i( -\pi/2)}.
	\end{equation}
	For a given triad, the more energy that appears in the convectively interacting components, $k',k''$, the higher the value of $|B|$. And, most importantly, the biphase for the convective interaction is $\beta = -\pi/2$. Plugging these results into the definition of $R$ in \eqref{eq:Rb}, we see that the cosine weighting of the biphase in $R$ means that $R=0$ for pure convective scale interactions. In other words, $R$ cannot actually detect convective scale interactions in turbulence. 
	
	(Of course, $R$ could detect interactions with $\beta=0$ which can result from other dynamical systems that exhibit pure quadratic non-linearity, like classical AM. But because the NSE contains only a convective non-linearity and not a purely quadratic term, we do not expect interactions with $\beta = 0$ to be significant in turbulence. It is worth noting that signals produced by quadratic AM or by a convective non-linearity are almost completely indistinguishable by visual inspection, although the difference in biphase is profound.) 
	
\begin{figure}
		\centering
		\begin{overpic}[width=0.80\textwidth]{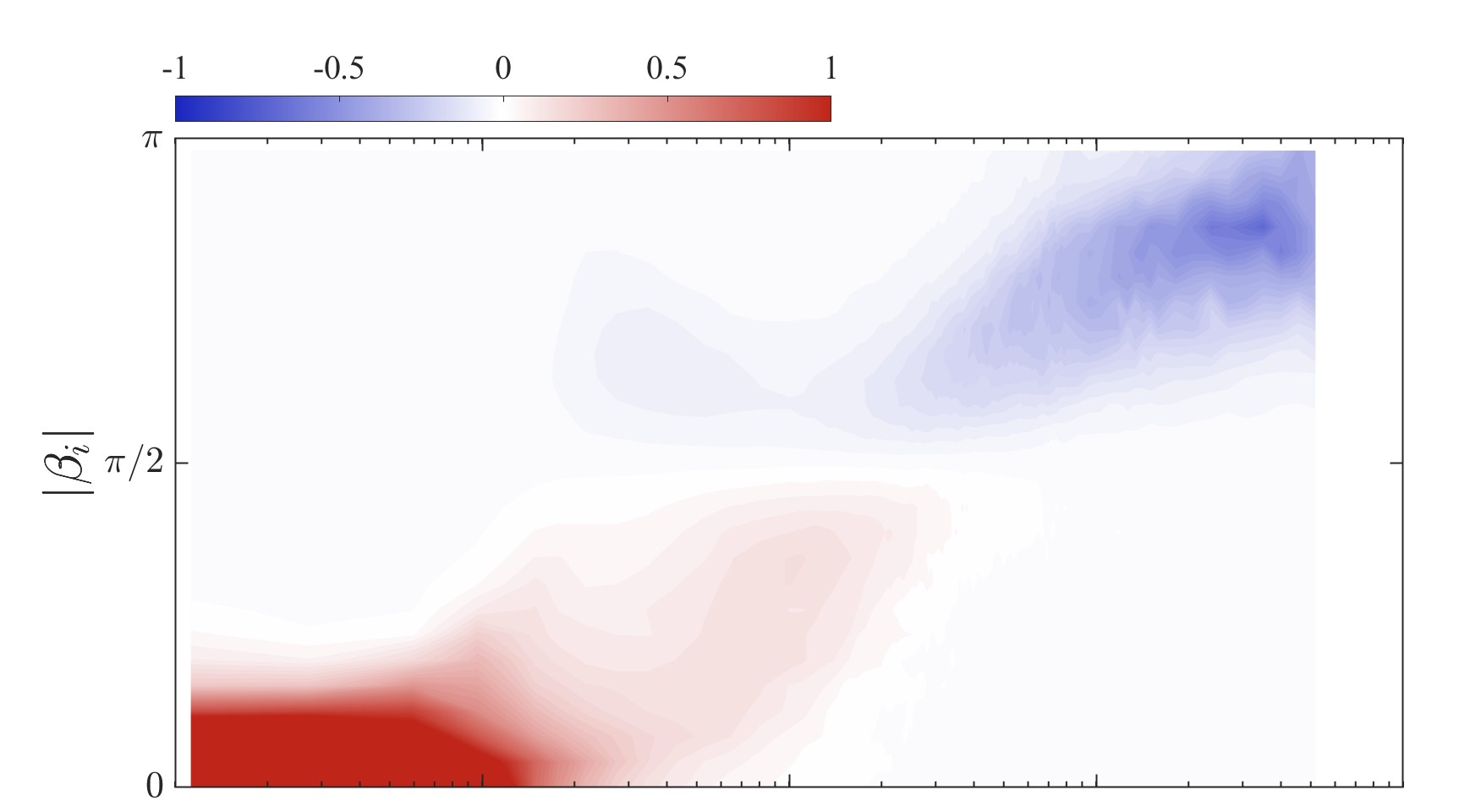}
			\put(0, 46){(a)}	
			\put(59, 47.5){$ \displaystyle \Delta R/\Delta \beta $} 	
		\end{overpic}
		\begin{overpic}[width=0.80\textwidth]{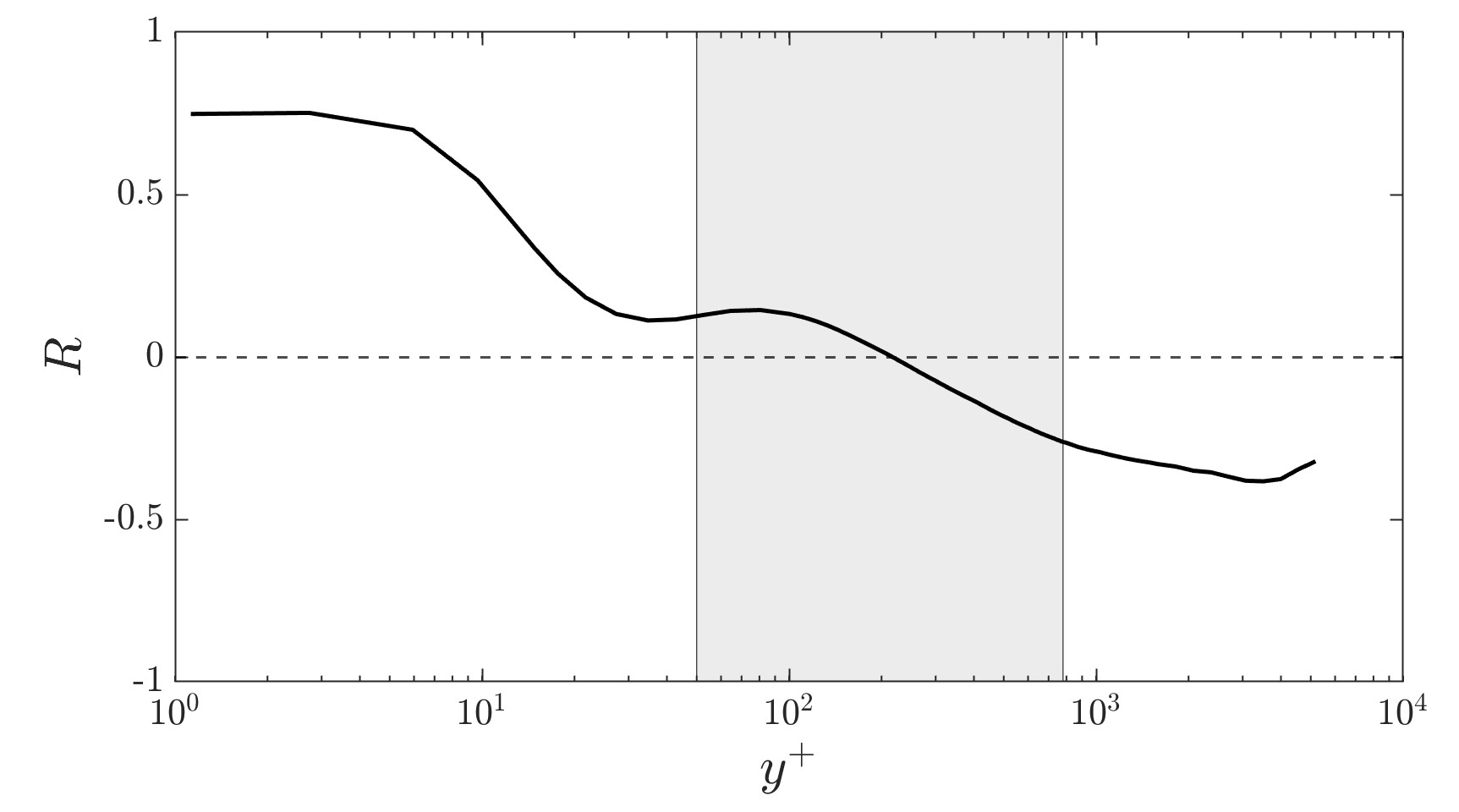}
			\put(0, 53){(b)}			
		\end{overpic}
		\caption{(a) The $R$-density, $\Delta R/\Delta \beta$, with respect to discrete biphase bins, $\beta_i$, according to \eqref{eq:Rdensity}, with $ \Delta \beta= 0.04 \pi$. The map was calculated from ensemble-averaging $84480$ streamwise/wall-normal snapshots of channel flow DNS data from \cite{lee2015direct}. (b) The classical $R$ profile is the integral of the $R$-density over all biphase bins. The gray region denotes the logarithmic layer. }
		\label{fig:map-beta-y-R} 
	\end{figure}

	Despite the model implication that $R=0$ for turbulence, the actual reported values of $R$ are not equal to zero across most of the wall-region. If convective scale interactions cannot contribute to $R$, what physical processes are then responsible for its non-zero value? To answer this, we can use the summation definition of $R$ in \eqref{eq:Rb} to decompose $R$ into contributions from different values of biphase  $\beta$ (irrespective of wavenumber) by binning the biphase into discrete bins denoted $\beta_i$, with uniform width $\Delta \beta$, according to:
		\begin{equation}
	R(y) =  \sum_{\beta_i} \overbrace{ \frac{2}{\Omega} \frac{1}{\Delta \beta}\;\sum_{\mathclap{\substack{ \forall \beta \, | \\ \beta_i < |\beta| \leq \beta_i + \Delta \beta}}}\; |B(\beta)| \cos{(\beta)}}^\text{$\Delta R/\Delta \beta$}\Delta \beta \label{eq:Rdensity}
	\end{equation}
	where the quantity in the brace is the $R$-density with respect to biphase, $\Delta R/\Delta \beta$. The $R$-density map was calculated from streamwise velocity fields from a direct numerical simulation (DNS) of a turbulent channel at $\text{Re}_\tau = 5200$ by \cite{lee2015direct}, with non-dimensional filter cutoff $k_f = 2\pi$ (non-dimensionalized by the channel half-height) and is shown in figure \ref{fig:map-beta-y-R}(a). The integral of the $R$-density yields the classical profile of $R$ shown in figure \ref{fig:map-beta-y-R}(b). 
	
	As expected, there is no contribution to $R$ from convective scale interactions with $\beta = \pm \pi/2$, due to the weighting. But there appears to be a significant positive contribution from triads with $\beta \approx 0$ in the viscous sublayer, and then a smaller negative contribution from triads with $3\pi/4 < |\beta| < \pi$ far from the wall. If these contributions are not associated with pure convective interactions, what do they represent? To answer this, we can return to the simplified model but now consider the effect of the mean flow interaction.
	
	\subsection{Model $R$ in the Presence of Mean Convection}
	
	Consider the model problem bispectrum when the interaction between the large-scale and the mean is dominant. The biphase $\beta =  -\pi/2 + \phi_0$ and $\phi_0 \to +\pi/2$ near the wall, where the large-scale convect faster than the local mean; and $\phi_0 \to -\pi/2$ far from the wall, where the large-scales convect slower than the mean. In other words, near the wall, we expect the strong mean interactions to shift the biphase toward zero, and away from the wall we expect the mean interactions to shift the biphase to $\pm \pi$, which is exactly what we observe in figure \ref{fig:map-beta-y-R}(a). The particularly high intensity of the $\beta = 0$ contributions in the viscous sublayer may also be a result of amplification by viscous effects, noted above. 
	
	Ultimately, because there is no pure AM in the NSE, the spectral decomposition of $R$ suggests that what $R$ really detects is interactions between large-scale features and the mean flow, and thus measures the difference between the velocity of the LSMs and the local mean velocity. The $\beta$-decomposed map of $R$ combined with the simplified model also allows us to consider a new interpretation for the zero-crossing location of $R$ that was somewhat unclear in previous analyses. $R$ crosses zero when the $R$-density is anti-symmetric about $\pi/2$, which occurs when the convective velocities of the large scales are distributed symmetrically about the local mean. By contrast, when the distribution of convection velocities is skewed toward velocities slower than the local mean, in the outer flow, then $\beta$ is skewed towards $\pm \pi$, and $R$ becomes negative. And, when the distribution of convection velocities is skewed toward velocities higher than the local mean, in the inner flow, then $\beta$ is skewed towards $0$, and $R$ becomes positive. So we suggest that the zero-crossing location of $R$ in the middle of the log layer indicates that the dominant LSMs advect at the local mean velocity in this location, consistent with the proposal of \cite{Chung2010}.
	
	But, since $R$ is inherently prevented from detecting convective scale interactions due to its cosine weighting of the biphase, we must define a new diagnostic that is weighted by $\sin{(\beta)}$ if we want to measure the relative importance of $\beta = \pm \pi/2$ interactions, quantitatively.

	\section{A Coefficient Designed for Detecting Convective Scale Interactions}
	
	\subsection{Definition and Spectral Decomposition of $R_T$}
	
	In order to incorporate a $\sin{(\beta)}$ weighting in the scale interaction analysis, we need to shift one of the two signals in the $R$ cross-correlation defined in \eqref{eq:R} by $\pi/2$ in phase. The simplest way to do this, assuming the spatial signals can be decomposed in a complex Fourier basis, is to differentiate one signal with respect to $x$. We apply this differentiation to the large-scale signal and define a new correlation coefficient $R_T$ following the same format as $R$:
	\begin{equation}
	R_T(y) = \frac{\left\langle \frac{\partial u_L}{\partial x} \mathcal{E}(u_S) \right\rangle }{\sqrt{\left\langle \left({\partial u_L}/{\partial x}\right)^2 \right\rangle} \sqrt{\left\langle \mathcal{E}(u_S)^2 \right\rangle \vphantom{\left\langle \left({\partial u_L}/{\partial x}\right)^2 \right\rangle}}}
	\label{eq:RT}
	\end{equation}
	As before, we rewrite this coefficient in spectral form to obtain:	
	\begin{align}
	R_T(y) = \frac{1}{\Omega_T}\sum_{\substack{\forall k \, | \\ \left| k\right|<k_f  }} \qquad  \sum_{\mathclap{ \substack{\forall k', k'' \, |  \\ \left| k' \right|, \left| k'' \right|>k_f \\ k'+k''=-k }}} (k'+k'') \left| B(k', k'') \right| \sin{[\beta(k', k'')]}  \label{eq:RTb}
	\end{align}
	where the new normalization factor is defined as $\Omega_T = \sqrt{\left\langle \left({\partial u_L  (x)}/{\partial x} \right)^2 \right\rangle } \; \sqrt{ \vphantom{\left\langle \left({\partial u_L  (x)}/{\partial x} \right)^2 \right\rangle } \left\langle 	\mathcal{E}{u_S}(x)^2 \right\rangle }$. (Using the derivative of the large-scale signal to better correlate with small scale activity was also suggested by \cite{Chung2010} although for different reasons.)
	
	Contrasting $R_T$ with the definition of $R$ in \eqref{eq:Rb}, we see that the new coefficient involves a similar summation of bispectral magnitude over the region of wavenumbers for scale interactions, but it is weighted by the sine of the biphase, instead of the cosine, and thus it is weighted towards capturing convective non-linear interactions with biphase $\beta = \pm \pi/2$. We can confirm this by decomposing $R_T$ with respect to $\beta$, like we did for $R$ in \eqref{eq:Rdensity}, as: 
\begin{equation}
R_T(y) =  \sum_{\beta_i} \overbrace{ \frac{2}{\Omega_T} \frac{1}{\Delta \beta}\;\sum_{\mathclap{\substack{ \forall \beta \, | \\ \beta_i < |\beta| \leq \beta_i + \Delta \beta}}}\; (k'+k'') | \, B(\beta) |\, \sin{(\beta)} }^\text{$\Delta R_T/\Delta \beta$}\Delta \beta \label{eq:RTdensity}
\end{equation}
	
\begin{figure}
		\centering
		\begin{overpic}[width=0.80\textwidth]{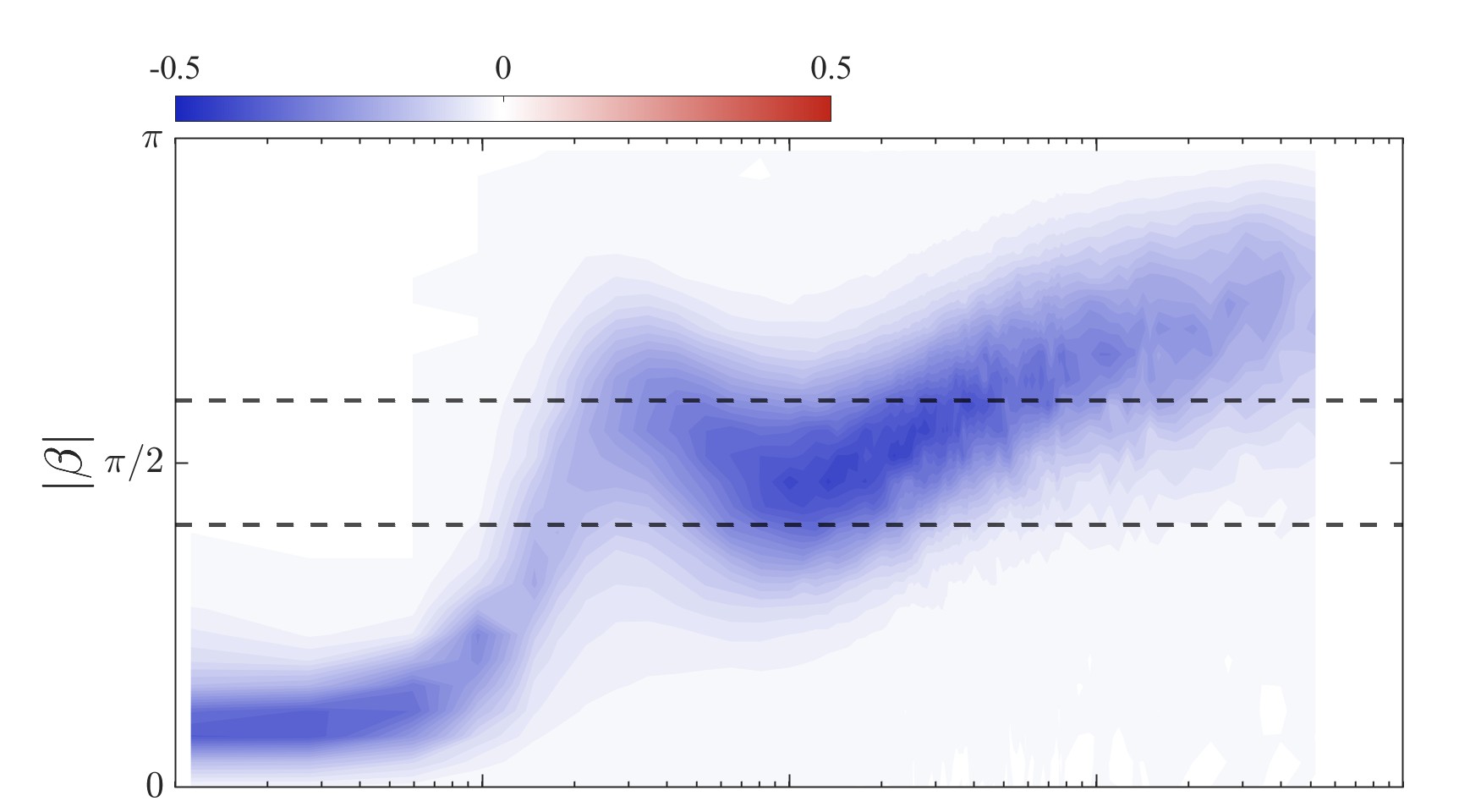}
			\put(0, 46){(a)}	
			\put(59, 47.5){$\Delta R_T/\Delta \beta$}
		\end{overpic}
		\begin{overpic}[width=0.80\textwidth]{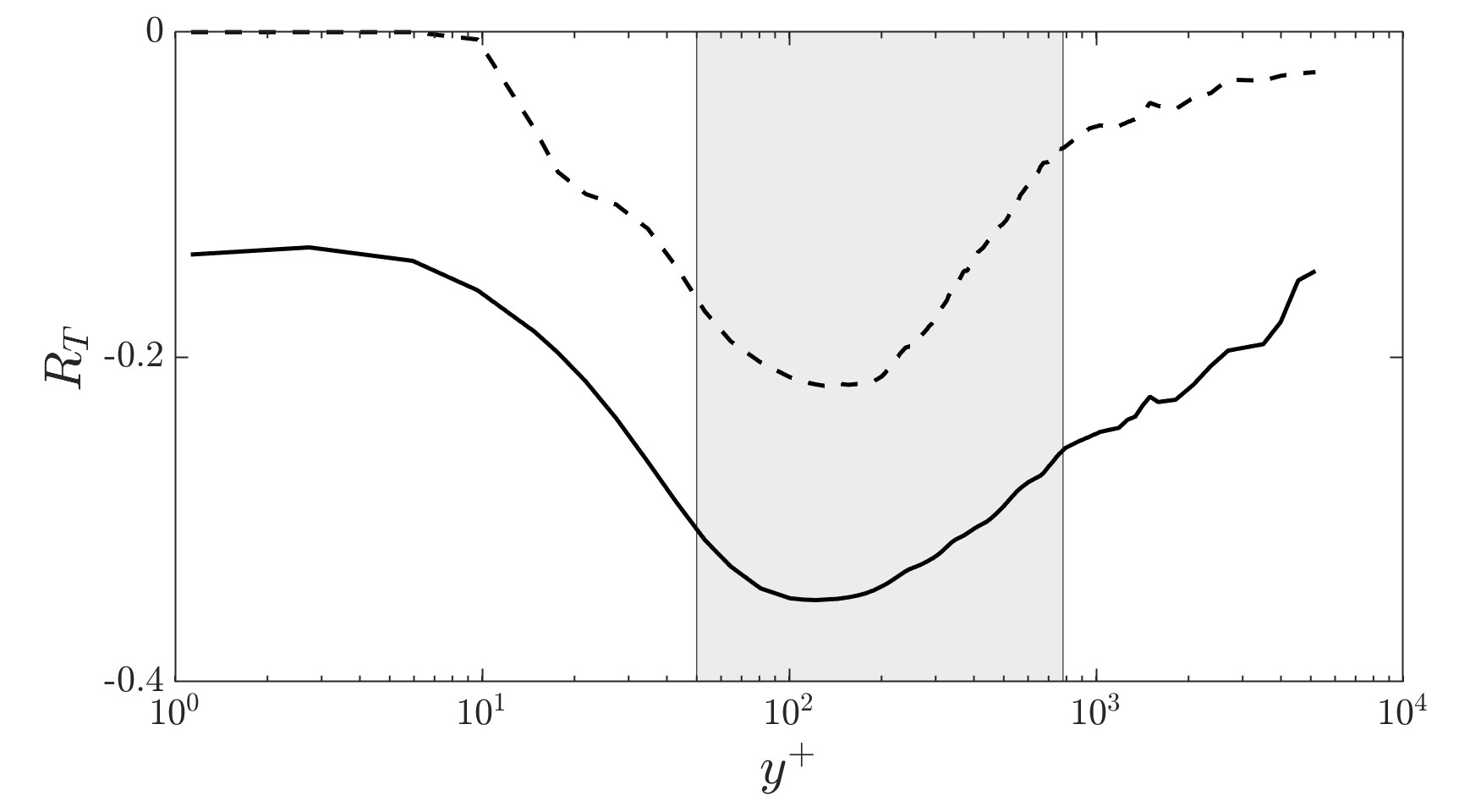}
			\put(0, 53){(b)}			
		\end{overpic}
		\caption{(a) The $R_T$-density, $\Delta R_T/\Delta \beta$, with respect to discrete biphase bins, $\beta_i$, according to \eqref{eq:RTdensity}. The dominant $R_T$ density for $|\beta| \approx \pi/2$ occurs in the log and buffer layers.  (b) The $R_T$ profile (solid line) is the integral of the $R_T$-density over all biphase bins, and can be calculated directly from the cross-correlation in \eqref{eq:RT}, using simple finite differences for evaluating the derivative of the filtered signal. The partial $R_T$ profile (dashed line) is the integral of the $R_T$ density between the two dashed lines in (a) at $|\beta| = \pi/2 \pm 0.3$. 
			  }
		\label{fig:map-beta-y-R2} 
	\end{figure}

	Figure \ref{fig:map-beta-y-R2}(a) shows the density $\Delta R_T/\Delta \beta$ and figure \ref{fig:map-beta-y-R2}(b) shows the profile of the new coefficient across the channel. We see that the contribution to $R_T$ from non-linear convective triads with biphase $\beta = \pm \pi/2$ is not suppressed; in fact, it seems to be the dominant contribution in the buffer and log layers, and that is where the profile of $R_T$ also reaches its maximum amplitude. By isolating a narrow region of $\beta$ around $\pm \pi/2$, we construct a profile of the contribution to $R_T$ from only these convective scale interactions, and compare that with the total $R_T$ obtained via cross-correlation. Both profiles appear nearly identical in shape, except for a translation in magnitude, which means that the $R_T$ profile obtained from simple cross-correlation captures the relative distribution of convective scale interactions across the channel, without the need for performing the tedious bispectral summation.  
	
The $R_T$ coefficient therefore provides a tool for comparing the relative strength of convective, non-linear interactions between large- and small-scales across turbulent  wall-bounded flows. Unlike the $R$ coefficient, $R_T$ does not suppress convective interactions. The location of the peak $R_T$ amplitude occurs in the log-layer, indicating that inter-scale interactions are most dominant there. The choice of filter cut-off tends to shift this location slightly: as $k_f$ decreases, the large-scale signal concentrates on even larger scales which are presumably centered farther from the wall, and thus the peak amplitude of $R_T$ shifts away from the wall. But the qualitative shape of the profile is relatively robust to the choice of filter cutoff, as was true for $R$ in \cite{Mathis2009}.

Because $R_T$ also depends on $\beta$, it too can be affected by mean interactions. However, unlike $R$, where the mean interactions induced spurious evidence for non-linearity, for $R_T$ the mean interactions merely suppress some of the evidence for true convective non-linearity, by reducing the biphase away from $\pm \pi/2$. However, because $R_T$ is weighted against the resulting $\beta=0$ quadratic non-linearities, this suppression should have a minimal effect on the shape of the $R_T$ profile.

\subsection{The Relationship Between $R_T$ and TKE Transport}

The reason that the location of maximal scale interactions, as detected by $R_T$, appears in the lower part of the log layer can be explained in terms of turbulent kinetic energy (TKE) transport across the near wall region. The wall-parallel form of the turbulent spectral transport, $\hat{T}$ is related directly to $R_T$. Neglecting wall-normal gradients, the turbulent spectral transport, is given by $\hat{T}(k_x,k_z;y) = -\frac{1}{2} \left\langle \hat{u}_i^* \widehat{\tfrac{\partial u_i u_j}{\partial x_j}} + \hat{u}_i \widehat{\tfrac{\partial u_i u_j}{\partial x_j}}^* \right\rangle$ where $i,j$ are indices for the streamwise ($x$) and spanwise ($z$) coordinates. Simplifying for the case of unidirectional ($i,j=x,k=k_x$) flow, and substituting for the complex Fourier modes defined in \eqref{eq:fourier}, we rewrite the transport in terms of the bispectrum as:
\begin{align}
\hat{T}(k;y) & = \sum_{\mathclap{ \substack{\forall k', k'' \, |\\ k'+k'' = -k }}}  (k'+k'') \, |B(k',k'')| \sin{\left[\beta(k',k'') \right]}  
\end{align}
and we see that $R_T$ is just a sum of a high-pass filtered version of the transport, denoted $\hat{T}_f$:
 	\begin{align}
 	R_T(y) = \frac{1}{\Omega_T}\sum_{\substack{\forall k \, | \\ \left| k\right|<k_f  }} \hat{T}_f(k;y), \quad \hat{T}_f(k;y) = \sum_{\mathclap{ \substack{\forall k', k'' \, |  \\ k'+k''=-k \\ \left| k' \right|, \left| k'' \right|>k_f}}} (k'+k'') \left| B(k', k'') \right| \sin(\beta(k', k'')). \label{eq:That}
\end{align}
Empirically, we find that $\beta(k',k'') > 0$ for $(k'+k'')<0$, and therefore $\hat{T}_f < 0$ for $k > 0$. By symmetry of the bispectrum, it follows that $\hat{T}_f < 0$ also for $k < 0$, and thus we observe that the transport is negative for all triads, i.e. the transport is always in the direction of the classical energy cascade, from large-scales $k$ to the small scales $k',k''$. And this corresponds to the $R_T$ profile being negative across the channel.

The fact that the profile of $R_T$ shows a maximum amplitude in the buffer and log layers is likely a consequence of the intense turbulent transport in these regions. Therefore, this new scale interaction coefficient, $R_T$, provides a simple way of examining the relative distribution of TKE transport associated with inter-scale energy exchange. 

\section{Conclusions}

The AM coefficient, $R$, cannot measure convective-type, non-linear interactions between different scales, and therefore should not be interpreted as a measure of interactions between large- and small-scale motions in wall-bounded turbulence. Based on a biphase decomposition of $R$ and a simple model of triadic scale interactions, we suggest that $R$ is really measuring linear interactions between LSMs and the mean, and is therefore a metric for local deviations from Taylor's hypothesis for LSMs. In place of $R$, we proposed a new coefficient $R_T$ that is weighted to appropriately capture convective scale interactions, and we showed how it can be interpreted naturally as a measure of turbulent TKE transport between large- and small-scale motions, which was found to be consistent with the classical energy cascade.

	\begin{acknowledgments} 
		The authors gratefully acknowledge the support of Israel Science Foundation grant 219/21. Declaration of Interests: the authors report no conflict of interest.
	\end{acknowledgments}
	
	\bibliographystyle{jfm}
	\bibliography{main_intro}

\end{document}